\documentclass[%
a4paper,							
11pt,								
bibliography=totoc,						
abstracton,					
]
{scrartcl}

\usepackage[a4paper,left=3.0cm,right=2.5cm, top=2.5cm, bottom=3.0cm]{geometry}
\usepackage[headsepline=.4pt,footsepline=.4pt,automark,autooneside=false,]{scrlayer-scrpage}
\clearscrheadfoot 
\automark[subsection]{section} 
\automark*[subsubsection]{subsection}
\ihead{\scriptsize\rightmark} 
\usepackage[english]{babel}
\usepackage[T1]{fontenc}
\usepackage[utf8]{inputenc}
\usepackage{lmodern} 
\usepackage[onehalfspacing]{setspace} 
\usepackage{eurosym}

\usepackage{xspace}
\usepackage{calc}

\usepackage{amsmath}
\usepackage{amssymb}
\usepackage{amsthm,thmtools}

\usepackage{mathtools}
\usepackage{bbm}
\begingroup
    \makeatletter
    \@for\theoremstyle:=definition,remark,plain\do{%
        \expandafter\g@addto@macro\csname th@\theoremstyle\endcsname{%
            \addtolength\thm@preskip\parskip
            }%
        }
\endgroup

\usepackage{chngcntr}
\counterwithin*{equation}{section}

\theoremstyle{plain}
\newtheorem{theorem}{Theorem}[section]

\theoremstyle{definition}

\newtheorem{problem}[theorem]{Problem}
\theoremstyle{remark}
\newtheorem{remark}[theorem]{Remark}
\theoremstyle{plain}
\newtheorem{proposition}[theorem]{Proposition}
\theoremstyle{plain}

\theoremstyle{remark}

\usepackage{tabularx}
\usepackage{booktabs}
\usepackage{multirow}
\usepackage{multicol}
\usepackage{pdflscape}

\usepackage{diagbox}

\usepackage[flushleft]{paralist}
\setdefaultenum{(i)}{(a)}{(1)}{(aa)}

\usepackage{color}

\usepackage{graphicx}
\usepackage{tikz}
\usetikzlibrary{calc,spy,backgrounds}
\usepackage{todonotes}
\usepackage[final]{showkeys}

\usepackage{fancyvrb}

\usepackage{float}
\usepackage[section]{placeins}
\usepackage[format=plain]{caption}

\usepackage{titleref}
\usepackage{hyperref}
\usepackage{cleveref}
\hypersetup{
    colorlinks,
		linkcolor={red},
    citecolor={blue},
    urlcolor={blue}
}

\newcommand{\LGD}{\mathrm{LGD}}

\newcommand{\aPoint}{A}
\newcommand{\dPoint}{D}

\newcommand{\R}{\mathbb{R}}

\newcommand{\Q}{\mathbb{Q}}
\newcommand{\N}{\mathbb{N}}
\newcommand{\1}{\mathbbm{1}}

\newcommand{\abs}[1]{\left|#1\right|}

\newcommand{\loss}{\mathrm{loss}}

\newcolumntype{C}{>{\centering\arraybackslash}X}

\newcommand{\CPU}{Intel(R) Core(TM) i7-8750H CPU @ 2.20\,GHz\xspace}
\newcommand{\RAM}{2x32\,GB (Dual Channel) Samsung SODIMM DDR4 RAM @ 2667 MHz\xspace}
\newcommand{\GPU}{NVIDIA GeForce RTX 2070 with Max-Q Design (8\,GB GDDR6 RAM)\xspace}
\newcommand{\OS}{Windows 10 Pro\xspace}

\SaveVerb{python}=(Intel-)Python 3.9=
\newcommand{\python}{\protect\UseVerb{python}\xspace}
\SaveVerb{tensorflow}=Tensorflow 2.8.0=
\newcommand{\tensorflow}{\protect\UseVerb{tensorflow}\xspace}
\SaveVerb{matlab}=Matlab 2022b=
\newcommand{\matlab}{\protect\UseVerb{matlab}\xspace}
\SaveVerb{ga}=ga=

\SaveVerb{fmincon}=fmincon=

\SaveVerb{fitdist}=fitdist=

\SaveVerb{lsqnonlin}=lsqnonlin=
\newcommand{\lsqnonlin}{\protect\UseVerb{lsqnonlin}\xspace}
\newcommand{\matlabGOtoolbox}{(Global) Optimization Toolbox\xspace}

\newcommand{\dateA}{26/09/2022\xspace}
\newcommand{\dateB}{05/12/2022\xspace}
\title{CDO calibration via Magnus Expansion and Deep Learning}
\author{Marco Di Francesco\thanks{UnipolSai Assicurazioni, via Stalingrado 45, Bologna, Italy, \textbf{e-mail}: marco.difrancesco@unipolsai.com}\and Kevin Kamm\thanks{Department of Mathematics and Statistics, 
Ume\aa\ University, Sweden.
\textbf{e-mail}: kevin.kamm@umu.se}}

\begin{document}
\thispagestyle{empty}\pagenumbering{roman}
\maketitle
\renewcommand{\thefootnote}{\Roman{footnote}}
\footnotetext[0]{The views expressed in this note are the only responsibility of the author and do
not represent in any way those of author's current employer. All errors are the only
responsibility of the author.}
\renewcommand{\thefootnote}{\arabic{footnote}}
\begin{abstract}
	In this paper, we improve the performance of the large basket approximation developed by \cite{Reisinger2011,Bush2011Arxiv,Reisinger2012} to calibrate Collateralized Debt Obligations (CDO) to iTraxx market data. The iTraxx tranches and index are computed using a basket of size $K=125$. In the context of the large basket approximation, it is assumed that this is sufficiently large to approximate it by a limit SPDE describing the portfolio loss of a basket with size $K\rightarrow\infty$. 
	
	For the resulting SPDE, we show four different numerical methods and demonstrate how the Magnus expansion can be applied to efficiently solve the large basket SPDE with high accuracy.
	
	Moreover, we will calibrate a structural model to the available market data. For this,
	it is important to efficiently infer the so-called initial \emph{distances to default} from the Credit Default Swap (CDS) quotes of the constituents of the iTraxx for the large basket approximation.
	We will show how Deep Learning techniques can help us to improve the performance of this step significantly.
	
	We will see in the end a good fit to the market data and develop a highly parallelizable numerical scheme using GPU and multithreading techniques.
\end{abstract}
\textbf{Keywords:} 
CDO, CDS, Risk-Neutral Pricing, Deep Learning, Magnus Expansion, SPDE\\\noindent
\textbf{Code availability:}
The code and data sets to produce the numerical experiments are available at
\url{https://github.com/kevinkamm/DeepCDO}.
\newpage
\pagestyle{scrheadings}\ihead{\scriptsize\rightmark}\pagenumbering{arabic}
\section{Introduction}\label{sec:introduction}
In this paper, we will demonstrate how to improve the Collateralized Debt Obligation (CDO) calibration scheme proposed by \cite{Reisinger2011,Bush2011Arxiv,Reisinger2012} by using the Magnus expansion and Deep Learning techniques. 

\begin{figure}%
\centering
\includegraphics[width=\columnwidth]{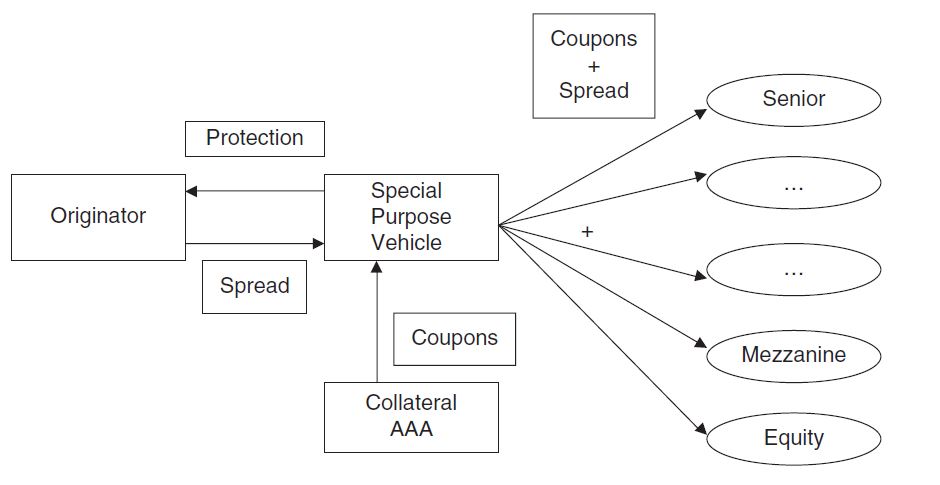}	
\caption{Overview of synthetic CDO market mechanism.}%
\label{fig:CDO}%
\end{figure}

CDOs in general are financial instruments related to the loss distribution of a pool of names.
In this paper, we will focus on so-called \emph{synthetic CDOs}. 
Alongside Figure \ref{fig:CDO}\footnote{This figure is taken from \cite[p.~214 Figure 8.2]{CherubiniDellaLunga}.}, let us explain the market mechanism of synthetic CDO: First, an intermediary, called \emph{Special Purpose Vehicle (SPV)} sells protection to a bank (Originator) by means of several Credit Default Swaps (CDS) and collects funds from investors by issuing securities. In this paper, the CDO will consist of $K=125$ different diverse CDS contracts. Due to CDS being worth zero initially, the funds raised by SPV are invested in default-free collateral. Afterwards, the revenue from interest payments and CDS spreads is passed back to the investors. 

The securities issued by the SPV are differentiated in a set called \emph{tranches}, defined by different degrees of seniority of debt. 
If some of the loans default and the money collected from the CDO is not enough to pay all investors, the SPV pays for the losses by reducing the amount of collateral and the principal of the tranches. Therefore, investors in the first tranche suffer losses first. After the first tranche has suffered its maximal loss, the next tranche is affected, and so on. For this reason, this mechanism is usually referred to as a \emph{waterfall} process.
The tranches are categorized by the percentage of the total portfolio losses. The first tranche, called \emph{equity tranche}, lies between $0\,\%$ and $3\,\%$ of the total losses, the second tranche (\emph{mezzanine tranche}) between $3\,\%$ and $6\,\%$ and is followed by senior and then super-senior tranches.

The purpose of a CDO is to sell a portfolio of assets to the general public. Moreover, it carries advantages for the seller and the buyer. For example, a bank may take advantages from securitization whereas an investor may take advantages from diversification. In fact, a CDO permits to invest in asset classes that otherwise would not have been permitted by usual regulations. Moreover, in times of low interest rates, CDOs have the potential to offer high yields. As happened in the last decade, an investment in CDO tranches represented a good opportunity due to low interest rates and high CDS spreads. 

However, CDOs were notoriously involved in the financial crisis in 2007. Without going into details, we would like to point the reader to the concept of X-Valuation Adjustments (XVA), which is dealing with derivative pricing while simultaneously taking default events into account. The interested reader is referred to \cite{Brigo2013} and more recently \cite{KM2022b}.
\paragraph*{Literature review.}
There exists a vast amount of literature concerning credit models and credit risk. For a comprehensive survey on those topics, we refer the reader to the monographs \cite{Rutkowsky2004}, 
\cite{Lando2004}, \cite{Duffie2012} and \cite{Schonbucher2003}.
As for CDO calibration, we recognize two different streams of models in the literature and refer the reader to \cite{Hu2015} and the references therein for a more detailed literature review. 

On the one hand, intensity-based models are typically used by practitioners to estimate the term structure of default probabilities of CDS quoted in the market under the risk-neutral measure, using the so-called bootstrapping technique. In these cases, a random default time is modeled directly, typically as a first jump of a Poisson process. The main advantages are ease of implementation, parallelism with interest rate models, and easier calibration to CDS spreads. 


On the other hand, in structural models, the default event is linked to the notion of corporate insolvency, which makes the credit event in contrast to intensity-based models economically interpretable. Moreover, structural models are easier to use in situations where we also need to model equity variables and take correlations into account. 
Finally, the concept of distance to default, which measures the debtor's leverage relative to the volatility of the value of its assets, can be used to reflect creditworthiness.


Since this paper follows the results established in \cite{Reisinger2011,Bush2011Arxiv} in the context of structural models closely, let us first of all give a high-level summary.

In \cite{Bush2011Arxiv}, the authors derive the theoretical background for the \emph{large basket approximation} in a structural diffusion model, which is fundamental for this paper. The structural model is designed as the distance of default of a CDS contract, from which the time of default can be computed as the first time the structural model hits the default barrier zero.
In this paper, they consider a diffusion model. It is assumed that for each individual name in the portfolio the distance of default has a common drift, an individual Brownian motion and a common Brownian motion with constant correlation. These simple SDEs can be considered as dynamics of particles in a larger system. Now, the idea is to increase the number of particles to infinite. They show that this idea with the proper definitions of \emph{an empirical measure} leads to an SPDE for the density of the \emph{limit measure}. From the solution of this SPDE we can compute the portfolio losses of an CDO with infinitely many entities. 

In \cite{Reisinger2011}, they extend the model by allowing jump-diffusions for the individual distances of default and discuss the validity of the large basket approximation. They demonstrate that a basket size of $K=125$ is large enough to behave like the large basket approximation and discuss the calibration problem (see Problem \ref{prob:cal1}) to CDO tranches, indices and CDS quotes in greater detail. In fact, this is the paper we follow the closest in this paper and show how the Magnus expansion and Deep Learning in the diffusion case can improve the performance of the proposed calibration scheme in \cite{Reisinger2011}.

\bigskip
The paper is structured as follows: First, in Section \ref{sec:CDO} we recall the relevant definitions of CDOs. We divide this section further by explaining the available market data in Section \ref{sec:iTRAXX} and recall the large basket approximation in Section \ref{sec:largeBasketApprox}. This is followed by the numerical experiments in Section \ref{sec:numerics}. Subsequently, we explain how to use two Deep-Neural-Networks (DNN) to infer the initial distances of default from the CDS quotes of the constituents of the iTraxx in Section \ref{sec:dnn}. After this, in Section \ref{sec:calCDO}, we introduce four different numerical schemes to compute the tranche spreads and the CDO index using the large basket approximation. We will see that the deterministic Magnus expansion works best and use it in Section \ref{sec:calibration} to calibrate our structural model to the market data. This is followed by a conclusion and possibilities for future research in Section \ref{sec:conclusion}.

\section{Collateralized Debt Obligations}\label{sec:CDO}
Let us begin this section by briefly recalling the necessary definitions for the CDO calibration problem and refer the reader for a more detailed treatment of this topic to
\cite[pp.~726\,ff.]{Brigo2006} and \cite[pp.~3\,ff.]{Reisinger2011}.

Henceforth, we will assume that the interest rate $r$ is deterministic and constant. Moreover, we will fix one maturity $T>0$ and assume that the following financial products have quarterly resettlements $T_j\coloneqq \alpha \cdot j \in [0,T]$, $j=1,\dots,n$, $n\in\N$, leading to an annuity of $\alpha=0.25$. Additionally, we will assume that the loss given default $\LGD=0.6$ is fixed as well.

Let us first of all recall the definition of the rate $\mathrm{c}_{0,T}$ making an individual CDS contract fair today. This CDS quote is given by
\begin{align}
	\mathrm{c}_{0,T} =
	\frac{
		\LGD\sum_{j=1}^{n}{
			\exp\left(-r T_j\right)
			\mathbb{E}^\Q\left[
				\1_{T_j>\tau} - \1_{T_{j-1}>\tau}
			\right]
		}
	}{
		\alpha\sum_{j=1}^{n}{
			\exp\left(-r T_j\right)
			\mathbb{E}^\Q\left[
				\1_{\tau>T_j}
			\right]
		}
	}
	\label{eq:cdsQuote}
\end{align}
where $\tau$ denotes the time of default of this particular entity. 

As mentioned in the introduction, for (synthetic) CDOs we will consider an entire portfolio of $K\in\N$ different CDS. The number $K$ is usually referred to as the \emph{size of the basket} of the CDO.

Now, we can define the \emph{(normalized) total portfolio loss}
\begin{align}
	L_t = \LGD\ \frac{1}{K} \sum_{k=1}^{K}{\1_{\tau_k \leq t}}
	\label{eq:portfolioLoss}
\end{align}
by taking the mean of defaulted entities up until the time $t\in [0,T]$ multiplied by the loss given default.\footnote{The portfolio loss is normalized in the sense that we set the notional for the individual CDS to $\frac{1}{K}$. Therefore, $\frac{1}{K} \sum_{k=1}^{K}{\1_{\tau_k \leq t}}$ is a percentage of the basket's defaulted entities at time $t$.} The time of default of the $k$-th entity is now denoted by $\tau_k$. If no entity defaults, the loss is zero and if all entities default, we
lose exactly the value of the loss given default.

One key feature of CDOs are their so-called tranches indicating a certain range of defaults. This interval of default percentages will be denoted by $[\aPoint,\dPoint]$ 
with \emph{attachment point} $\aPoint$ (in percent) and \emph{detachment point} $\dPoint$ (in percent).
The so-called \emph{outstanding notional} for an $[\aPoint,\dPoint]$-tranche can then be defined by
\begin{align}
	Z_t = \left(\dPoint-L_t\right)^+ - \left(\aPoint-L_t\right)^+.
	\label{eq:outstandingTrancheNotional}
\end{align}

Similar to the CDS quotes \eqref{eq:cdsQuote}, the spread of a \emph{single-tranche CDO (STCDO)} can be computed by
\begin{align}
	\mathrm{C}_{0,T} = 
	\frac{
		\sum_{j=1}^{n}{
			\exp\left(-r T_{j}\right)
			\mathbb{E}^\Q\left[
				Z_{T_{j-1}}-Z_{T_{j}}
			\right]
		}
	}{
		\alpha
		\sum_{j=1}^{n}{
			\exp\left(-r T_{j}\right)
			\mathbb{E}^\Q\left[
				Z_{T_{j-1}}
			\right]
		}
	}.
	\label{eq:STCDOspread}
\end{align}

The market also quotes CDO indices without considering individual tranches. 
The \emph{outstanding index notional} is defined by
\begin{align}
	Z_t^I = \frac{1}{K} \sum_{k=1}^{K}{\1_{\tau_k > t}}
	\label{eq:oustandingIndexNotional}
\end{align}
and the corresponding \emph{CDO index} by
\begin{align}
	\mathrm{I}_{0,T} = 
	\frac{
		\LGD \sum_{j=1}^{n}{
			\exp\left(-r T_j\right)
			\mathbb{E}^{\Q}\left[
				Z_{T_{j-1}}^I - Z_{T_{j}}^I 
			\right]
		}
	}{
		\alpha \sum_{j=1}^{n}{
			\exp\left(-r T_j\right)
			\mathbb{E}^{\Q}\left[
				Z_{T_{j}}^I
			\right]
		}
	}.
	\label{eq:CDOindex}
\end{align}

\subsection{iTraxx Data}\label{sec:iTRAXX}
Standardized CDO tranche quotes are available in the market: for example iTraxx indices and CDX indices are synthetic unfunded CDO quoted in the market. 
In particular, iTraxx specializes in the European market, but indices are also available for the Australian and the Asian market. We focus on iTraxx Europe, the index constituted of the main $125$ equally weighted CDS on investment grade European corporate entities. The names are revised and the index rolled over every six months, in March and September. The standard maturities are $5$ and $10$ years.

The quote convention is that the prices are expressed in basis points referred to the CDS premium for the tranche rendering the contract fair at inception. Since the premium for the equity tranche and often for the mezzanine tranche is usually very large, the market practice is to pay those tranches as a fixed running premium of $100\,$bps plus an upfront payment, computed in a way such that the total value is zero at inception. 

For our calibration exercise we will use the maturity $T=5$ years and report iTraxx Europe market data at \dateB obtained from Bloomberg with the aforementioned adjustment for the relevant tranches and the index alongside the calibration results later in Table \ref{tab:calibrationError}.


We refer to the code for another experiment at \dateA with similar results.
\subsection{Large Basket Approximation}\label{sec:largeBasketApprox}
Let us briefly recall the necessary ingredients of the large basket approximation from \cite{Reisinger2011,Bush2011Arxiv}:

\begin{proposition}\label{prop:largeBasket}%
	Let the distances of default $X_t^k$ be given by
	\begin{align}
		dX_t^k = \beta\ dt + \sqrt{1-\rho}\ dW_t^k + \sqrt{\rho}\ dM_t, \quad \beta=\frac{r-\frac{1}{2}\sigma^2}{\sigma},
		\label{eq:distanceToDefault}
	\end{align}
	where $W_t^k$ and $M_t$ are independent standard Brownian motions for $k=1,\dots,K$, $K\in\N$. Now, define the \emph{empirical measure} 
	\begin{align*}
		\nu_{K,t}\coloneqq \frac{1}{K}\ \sum_{k=1}^{K}{\delta_{X_t^k}},
	\end{align*}
	where $\delta_x$ denotes the Dirac-delta, which is one if $x=0$ and otherwise zero.
	
	If $\mathbb{E}^{\Q}\left[\left(X_0^k\right)^2\right]< \infty$ and $X_0^k$ are exchangeable, then the limit $\nu_t$ exists and its density $v$ is the solution to the SPDE
	\begin{align}
		d v(t,x) = 
			-\beta \left(\partial_x v\right)(t,x) dt + \frac{1}{2} \left(\partial_{xx} v\right)(t,x) dt - \sqrt{\rho} \left(\partial_{x} v\right)(t,x) dM_t.
		\label{eq:largeBasketSPDE}
	\end{align}
\end{proposition}
Now, for the application to CDO calibration, we need to monitor the default events $X_t^k=0$ and therefore introduce a default-barrier in the formulation of the SPDE. \cite{Reisinger2011} suggest to assume that the defaults can only be observed at a discrete set of times $\mathcal{T}^N\coloneqq\left\{T_1,\dots,T_N\right\}$, e.g., quarterly, which leads to a reformulation of Proposition \ref{prop:largeBasket} like follows:

We will assume that if a firm's value touches the default barrier, then it cannot recover and will be removed from the basket, i.e., define
\begin{align}
	\tau^k \coloneqq \inf_{t\in \mathcal{T}^N} X_t^k\leq 0
	\label{eq:defaultTimes}
\end{align}
and set
\begin{align*}
	X_t^k =0 \text{ on } t\geq \tau^k.
\end{align*}
Suppose, that $x_0^k$, $k=1,\dots,K$, are given. Then, for a Monte-Carlo implementation one can now simulate the $k$ different distances to default \eqref{eq:distanceToDefault} and compute the corresponding times of default \eqref{eq:defaultTimes}. This in turn can be inserted into the formulas \eqref{eq:cdsQuote}--\eqref{eq:CDOindex} and will serve as a sanity check in the numerical section.

This leads to the following large basket approximation (cf. \cite[p.~13 Proposition 4.2]{Reisinger2011}).
\begin{theorem}\label{thm:largeBasket}%
	Let everything be as in Proposition \ref{prop:largeBasket} for $X_t^k$ on $(0,\tau^k)$,
	then the limit empirical measure 
	\begin{align*}
		\bar{\nu}_t = \lim_{K\rightarrow \infty} \bar{\nu}_{K,t}
	\end{align*}
	exists and has the form $\bar{\nu}_t = \nu_t + \frac{\delta_0}{\LGD} L_t$, where 
	$\nu_t$ has a density $v$ satisfying \eqref{eq:largeBasketSPDE} for 
	$t\in \left(T_n,T_{n+1}\right)$, $n=0,\dots,N$, $N\in\N$, and 
	\begin{align*}
		L_t = \LGD \left(1-\int_{\R}^{}{v(t,x) dx}\right).
	\end{align*}
	Moreover, with probability one 
	\begin{align*}
		\lim_{t\downarrow T_n} v(t,x)=
		\begin{cases}
			\lim_{t\uparrow T_n} v(t,x), & x>0 \\ 
			0, & x\leq 0.\\ 
		\end{cases}
	\end{align*}
\end{theorem}

This leads to the following iterative scheme for the CDO evaluation using \eqref{eq:largeBasketSPDE}:
\begin{align}
	v(t,x)=
	\begin{cases}
		0, & x\leq 0,\quad t=T_{n+1}\\ 
		v^{(n)}(t-T_n,x),& t\in (T_n,T_{n+1}],\\ 
	\end{cases}
	\label{eq:schemeSPDE}
\end{align}
where
\begin{align*}
	d v^{(n)}(t,x) &= -\beta \left(\partial_{x} v^{(n)}\right)(t,x) dt + \frac{1}{2} \left(\partial_{xx} v^{(n)}\right)(t,x) dt - \sqrt{\rho} \left(\partial_{x} v^{(n)}\right)(t,x) dM_t,\\
	v^{(n)}\left(0,x\right) &= v^{(n-1)}\left(T_n,x\right)\1_{x> 0}.
\end{align*}

Alternatively, \cite{Reisinger2011} note that \eqref{eq:largeBasketSPDE} can be rewritten in terms of a deterministic PDE evaluated at points shifted by Brownian increments, which can be verified by applying Itô's formula. Therefore, we can consider the following problem
\begin{align}
	v(t,x)=
	\begin{cases}
		0, & x\leq 0,\quad t=T_{n+1}\\ 
		u^{(n)}(t-T_n,x-\sqrt{\rho}\left(M_t-M_{T_n}\right))& t\in (T_n,T_{n+1}],\\ 
	\end{cases}
	\label{eq:schemePDE}
\end{align}
where
\begin{align}
	\begin{aligned}[c]\arraycolsep=0pt
		\left(\partial_{t} u^{(n)}\right) &= \frac{1}{2}\left(1-\rho\right) \left(\partial_{xx} u^{(n)}\right) - \beta \left(\partial_{x} u^{(n)}\right)\\
		u^{(n)}\left(0,x\right) &= u^{(n-1)}\left(T_n,x\right)\1_{x> 0}.
	\end{aligned}
	\label{eq:largeBasketPDE}
\end{align}

\paragraph*{Initial datum $v^{(0)}(x)$.}
The initial density for the large basket approximation can be computed from the initial distances of default $x^k$ for the individual constituents of the iTraxx as follows (cf. \cite[pp.~18--19]{Reisinger2011}):
\begin{align*}
	v^{(0)}(x)=\frac{1}{K}\sum_{k=1}^{K}{\delta\left(x-x^k\right)}.
\end{align*}
They also note that this initial datum can lead to a poor accuracy, since it is very rough and a smoothed version leads to second order accuracy for finite-difference schemes.
Therefore, let us fix a homogeneous space grid 
$\mathbb{X}_{a,b}^d\coloneqq \left\{x_i\coloneqq a+ i \Delta x \colon \Delta x = \frac{b-a}{d+1},\ i=0,\dots,d+1\right\}$ and set
\begin{align*}
	\Phi_k(x) \coloneqq 
	\frac{1}{\Delta x} 
		\min\left(
			\max\left(x-x_i + \Delta x, 0\right),
			\max\left(x_i-x + \Delta x, 0\right)
		\right).
\end{align*}
The initial datum $v^{(0)}(x)$ evaluated in our space grid points $x_i$ is then given by
\begin{align}
	u_i^0 \coloneqq v_i^0 \coloneqq \frac{1}{\Delta x} \int_{a}^{b}{\Phi_k(x) v^{(0)}(x) dx} 
	\label{eq:initialDatum}
\end{align}

In Section \ref{sec:numerics}, we will apply the Euler-Maruyama scheme and the stochastic Magnus expansion to \eqref{eq:schemeSPDE}. Moreover, we will compare these to the theta-scheme proposed by \cite{Reisinger2011} and the deterministic Magnus expansion for the previous PDE \eqref{eq:schemePDE} with spline interpolation. For all schemes, we will use the smoothed version \eqref{eq:initialDatum} of the initial datum for given initial distances of default $x^k$.

\section{Numerical Experiments}\label{sec:numerics}
In this section, we will explain how to calibrate a structural model to the available CDO market data. For this, we will first recall the calibration problem from \cite[p.~22 Problem 1]{Reisinger2011}:

\begin{problem}\label{prob:cal1}%
Let the interest rate $r$ be constant and fixed. For given market spreads at $t=0$ of CDO tranches $\mathrm{C}_{0,T}^j$ and the CDO index $\mathrm{I}_{0,T}$ for maturity $T>0$ and
$[\aPoint_j,\dPoint_j]$-tranches $j=1,\dots,J$, $J\in\N$, and given spreads $\mathrm{c}_{0,T}\coloneqq \left(\mathrm{c}_{0,T}^1,\dots,\mathrm{c}_{0,T}^K\right)$ for CDS written on $K$ underlying companies solve the minimization problem
\begin{align}
	\min_{\rho\in [0,1),\sigma\in \Sigma}
		\sum_{j=1}^{J}{
			\left(
				\mathrm{C}_{0,T}^j -
				\mathrm{C}_{0,T}^j(\rho,\sigma,x_0)
			\right)^2
		}+
		\left(
			\mathrm{I}_{0,T} -
			\mathrm{I}_{0,T}(\rho,\sigma,x_0)
		\right)^2
	,
	\label{eq:calOuter}
\end{align}
subject to
\begin{align}
	\mathrm{c}_{0,T} = \mathrm{c}_{0,T}(\rho,\sigma,x_0)
	\label{eq:calInner},
\end{align}
where $\Sigma \subset \R_{>0}$ and $x_0\in\mathcal{X}\subset\R^K_{\geq 0}$ are suitable subsets.
\end{problem}
As we can see, Problem \ref{prob:cal1} consists of two nested minimization problems. The outer minimization \eqref{eq:calOuter} is dependent on the inner minimization \eqref{eq:calInner} by inferring the initial distances to default $x_0$ from the CDS quotes of the constituents of the iTraxx. The authors of \cite{Reisinger2011} suggest a Monte-Carlo approach with carefully selected starting points for the optimization problem \eqref{eq:calInner} and caution also that this is a computationally heavy task. 

Our idea is to disentangle the calibration problems. In Section \ref{sec:dnn}, we will demonstrate how to train a Deep-Neural-Network to learn $x_0$ for given parameters $\rho$ and $\sigma$, such that \eqref{eq:calInner} is satisfied. This removes the Monte-Carlo estimation from the entire calibration problem, since we will use the large basket approximation developed in \cite{Reisinger2011,Bush2011Arxiv,Reisinger2012} for the minimization \eqref{eq:calOuter}, for which we will use the Magnus expansion and is referred to Section \ref{sec:calCDO}. We will see that this new approach improves the overall performance of the calibration problem significantly.

For the calibration we used \matlab with the \matlabGOtoolbox
and for the DNN \python with \tensorflow
running on \OS, on a machine with the following specifications: processor
\CPU and \RAM, and a \GPU.

\subsection{Deep Neural Networks}\label{sec:dnn}
In this section, we will tackle \eqref{eq:calInner}. We will introduce two different DNNs to derive $x_0$ for given parameters $\rho,\sigma$, such that \eqref{eq:calInner} is satisfied, to be more precise we want to define two DNNs $f,g$, such that
\begin{align*}
	\mathrm{c}_{0,T} = g(\rho,\beta,x_0^f), \quad f(\rho,\beta)=x_0^f.
\end{align*}
The DNN\footnote{Sometimes we will write $f(\rho,\sigma)$ or $g(\rho,\sigma)$ instead of using $\beta$ as an input; in these cases $\beta$ can be first determined by using the fixed interest rate $r$ and the input $\sigma$. The advantage of using $\beta$ for the training of the DNNs is that for a large enough interval for $\beta$, it does not have to be re-trained for a small change in the interest rate.} $g$ will serve as an interpolation of CDS quotes for given parameters $\rho,\sigma,x_0$, while the DNN $f$ is supposed to find $x_0$, such that for all parameters $\rho,\sigma$ and fixed interest rate $r$, the market CDS quotes $\mathrm{c}_{0,T}$ are matched.

\paragraph*{Interpolation network for CDS quotes.}
Let us first of all train a DNN $g(\rho,\sigma,x_0)$ for predicting CDS quotes \eqref{eq:cdsQuote} from Monte-Carlo prices using a range of different parameters $\rho$ and $\sigma$ for the distances of default. As a reminder (cf. \cite[p.~5 Equation 3.4]{Reisinger2011}), the default event $\tau$ in our model is defined as
\begin{align}
	\tau \coloneqq \inf_{t>0}\left\{X_t\leq 0\right\}, 
	\label{eq:default}
\end{align}
where 
\begin{align*}
	dX_t = \beta\ dt + \sqrt{1-\rho}\ dW_t + \sqrt{\rho}\ dM_t,\ X_0=x_0 \in\R_{>0}, \quad \beta\coloneqq \frac{\left(r-\frac{1}{2}\sigma^2\right)}{\sigma},
\end{align*}
and $W_t$ is a standard $\Q$-Brownian motion.

We simulated $M=10^5$ paths of the independent Brownian motions $M_t$ and $W_t$ and generated a dataset of $2^{17}$ different tuples of $(\rho,\beta,x_0)$, each drawn from a uniform distribution. For $\rho$ we are limited to the range $[0,1)$, while we use for $\beta$ the range determined from $\sigma \in \left[0.01,0.5\right]$ and fixed $r$. Additionally, we found that $x_0\in (0,6)$ is sufficient for our purposes.

Using \eqref{eq:default}, we can simulate the corresponding Monte-Carlo prices by evaluating \eqref{eq:cdsQuote}.

With this training set at hand, we are in a position to train the DNN $g$ by minimizing the loss 
\begin{align*}
	\loss_g\left(\mathrm{c}^{\mathrm{MC}}_{0,T},\mathrm{c}^{g}_{0,T}\right) \coloneqq 
	\left(\sqrt{\mathrm{c}^{\mathrm{MC}}_{0,T}}-\sqrt{\mathrm{c}^{g}_{0,T}}\right)^2.
\end{align*}
With a range of $(0,6)$ for $x_0$ the values of the CDS quotes can get quite small and are below 1 in our experiments. Therefore, we decided to use square-roots in the loss function, to avoid that the DNN neglects small values and also to force positive values.

The design of the neural network is based on suggestions by \cite{He2022}, in that we use convolution layers instead of only fully connected layers. We tested both architectures and found the design with one-dimensional convolutional layers performed better. In Table \ref{tab:dnn1}, we show the detailed network design.

\begin{table}[t]%
\centering
\caption{Network design of the DNN $g$ for interpolating single CDS quotes.}
\begin{tabular}{*{5}{c}}
	Type & Filters & Kernel Size & Units & Activation\\
	\toprule
	Conv-1d & 16 & 4 & & Relu\\
	Conv-1d & 32 & 16 & & Relu\\
	Conv-1d & 64 & 32 & & Relu\\
	Conv-1d & 128 & 64 & & Relu\\
	Flatten & & & &\\
	Dense & & & 256 & Relu\\
	Dense & & & 128 & Relu\\
	Dense & & & 64 & Relu\\
	Dense & & & 8 & \\
\end{tabular}
\label{tab:dnn1}
\end{table}

In our experiments, we found that the accuracy of the DNN $g$ was much better with more than a single output and taking the average of the last layer, i.e.
\begin{align*}
	\mathrm{c}^{g}_{0,T}\left(\rho,\beta,x_0\right) \coloneqq \frac{1}{8} \sum_{i=1}^{8}{\left[g\left(\rho,\beta,x_0\right)\right]_i}.
\end{align*}

After training the DNN with Adam as optimizer using 30 epochs and a batch size of 128, we now have an interpolator of CDS quotes for the entire range of parameters.

\begin{figure}[b]%
\centering
\includegraphics[width=.5\columnwidth]{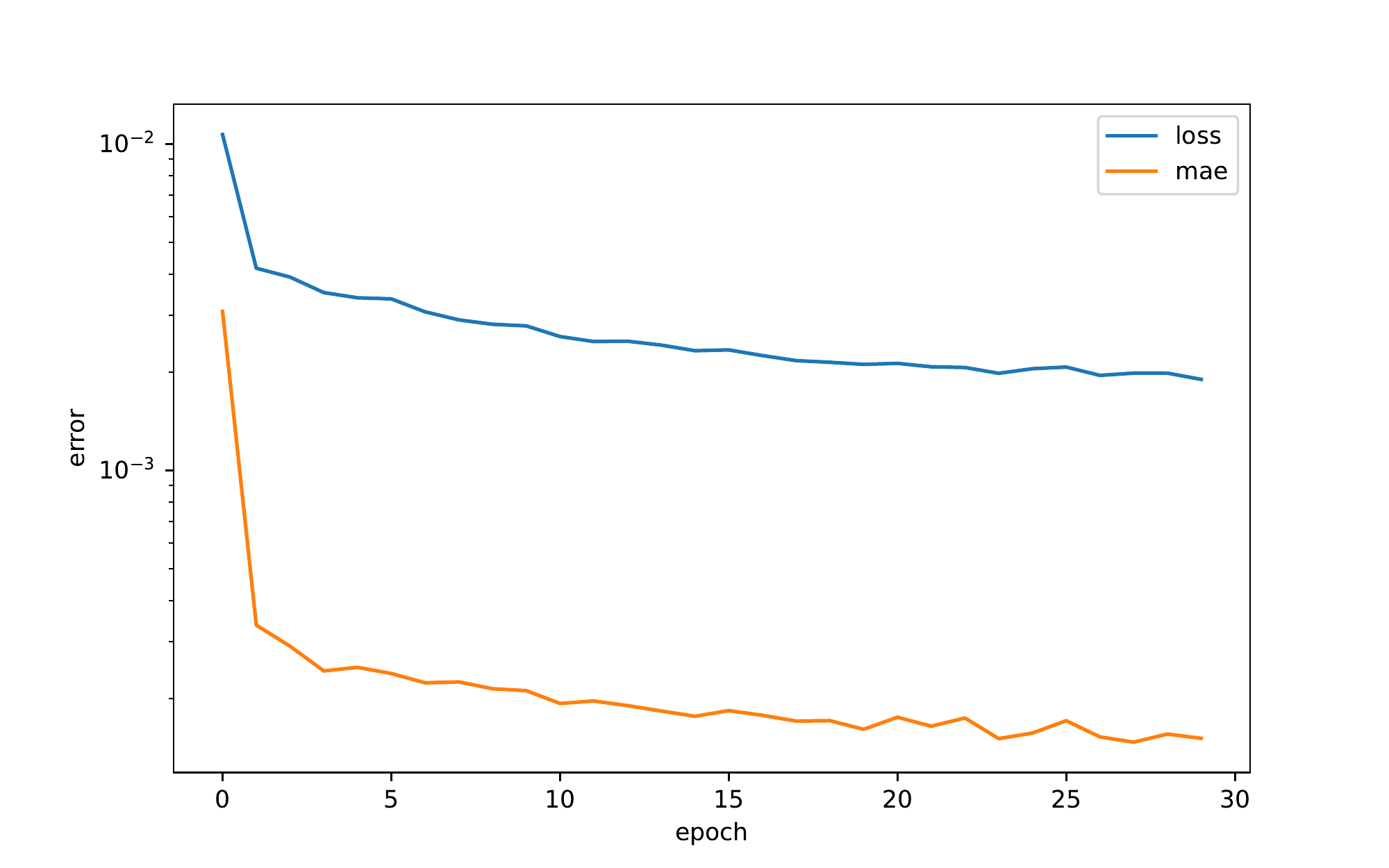}%
\caption{Training losses for $g$ at \dateB.}%
\label{fig:CDSLosses}%
\end{figure}

In Figure \ref{fig:CDSLosses}, we can see the training losses on the y-axis in a log-scale for each epoch (x-axis). 
The blue line corresponds to the loss function $\mathrm{loss}_g$ and the orange line is the mean absolute error of $\mathrm{c}^{g}_{0,T}$ and $\mathrm{c}_{0,T}$. We can see that both curves behave similarly. First, they rapidly decline, then their decrease slows down with a downwards trend till the last epoch.

\paragraph*{Network for finding the initial datum $x_0$.}
The aim is to find the initial distances to default with given parameters $\rho,\beta$, such that
the previously trained DNN $g$ is close to the market quotes for all the constituents of the iTraxx, i.e.
\begin{align*}
	\left[g(\rho,\beta,\left[f(\rho,\beta)\right]_k)\right]_{k=1,\dots,K}=
	\left[\mathrm{c}_{0,T}\right]_{k=1,\dots,K}.
\end{align*}
For this, we exploit two things as explained in \cite{Reisinger2011}: By exchangeablity we may sort the market CDS quotes. We will use a descending order. This will correspond for fixed $\rho,\beta$ to a vector $x_0\in \R^K_{\geq 0}$ with increasing values, because the closer $X_t$ is to zero at the start, the more likely it is to default and therefore the higher the CDS quote.

We chose the network design of $f$ very similar to Table \ref{tab:dnn1} and present it in 
Table \ref{tab:dnn2}.
\begin{table}[t]%
\centering
\caption{Network design of the DNN $f$ for inferring the initial distance to default $x_0$.}
\begin{tabular}{*{5}{c}}
	Type & Filters & Kernel Size & Units & Activation\\
	\toprule
	Conv-1d & 16 & 2 & & Relu\\
	Conv-1d & 32 & 16 & & Relu\\
	Conv-1d & 64 & 32 & & Relu\\
	Conv-1d & 128 & 64 & & Relu\\
	Flatten & & & &\\
	Dense & & & 4 K & Relu\\
	Dense & & & 2 K & Relu\\
	Dense & & & K & \\
\end{tabular}
\label{tab:dnn2}
\end{table}
We chose the loss function also similar to the previous network with the difference that we use in this case mean absolute percentages, i.e., we set
\begin{align*}
	\mathrm{c}^{f}_{0,T}\left(\rho,\beta\right)\coloneqq
	\mathrm{c}^{g}_{0,T}\left(\rho,\beta,f(\rho,\beta)\right),
\end{align*}
where the network for g was already trained beforehand and is now fixed, and define
\begin{align*}
	\loss_f\left(\mathrm{c}_{0,T},\mathrm{c}_{0,T}^f\right)\coloneqq
	100 \cdot \frac{\abs{\sqrt{\mathrm{c}_{0,T}} - \sqrt{\mathrm{c}_{0,T}^f}}}{\sqrt{\mathrm{c}_{0,T}}}.
\end{align*}
For the training set, we also use the same procedure as beforehand by drawing uniform random numbers from the aforementioned intervals for $\rho,\beta$, but this time the market CDS quotes $\mathrm{c}_{0,T}$ remain fixed. In total, we used $40$ epochs with $100$ batches of size $128$ of different parameters for training $f$ also using Adam as the optimizer.

\begin{figure}[b]%
\centering
\includegraphics[width=.5\columnwidth]{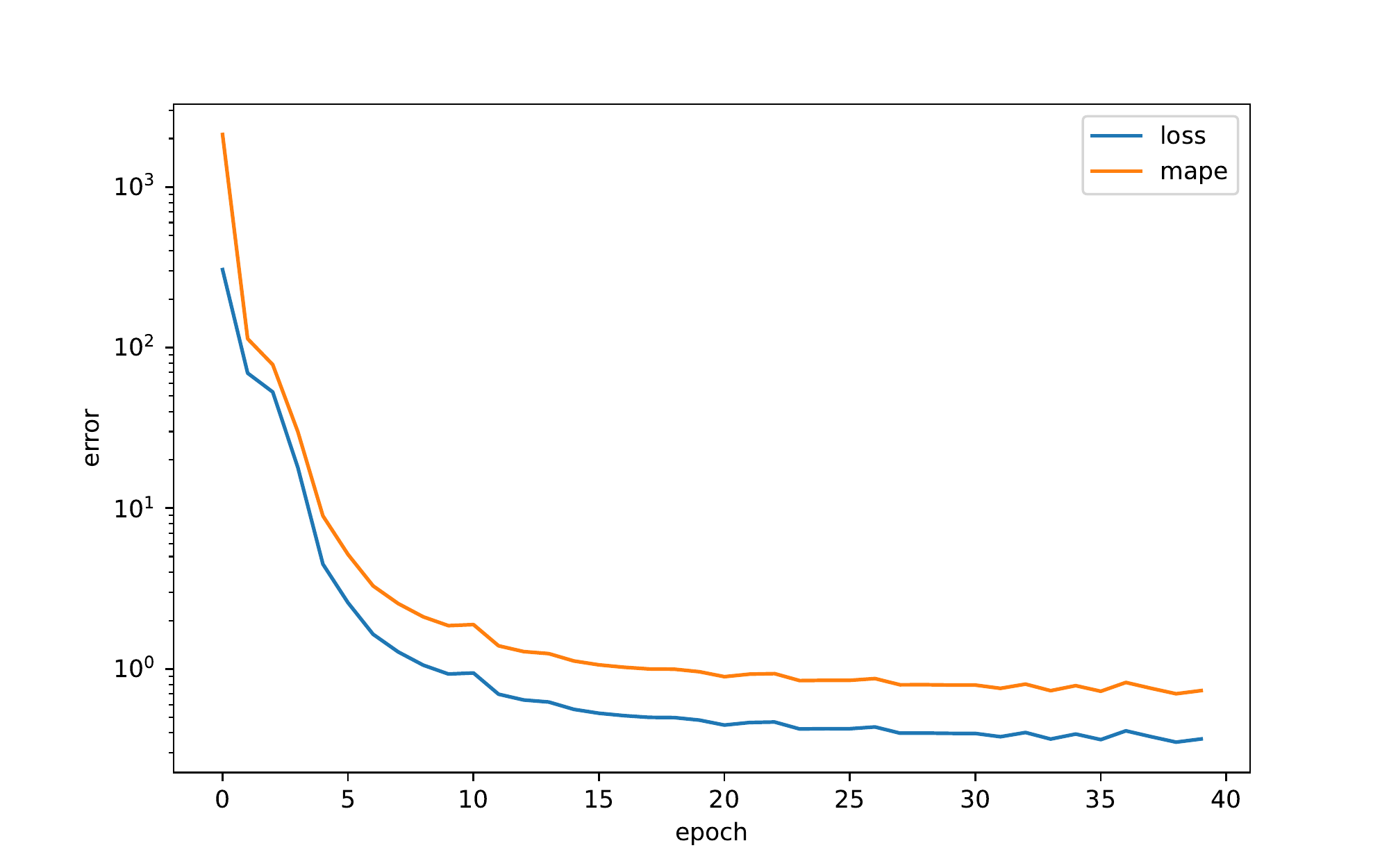}%
\caption{Training losses for $f$ at \dateB.}%
\label{fig:IniLosses}%
\end{figure}

In Figure \ref{fig:IniLosses}, we can see the training losses on the y-axis in a log-scale for each epoch (x-axis). 
The blue line corresponds to the loss function $\mathrm{loss}_f$ and the orange line is the mean absolute percentage error of $\mathrm{c}^{f}_{0,T}$ and $\mathrm{c}_{0,T}$. We can see that both curves behave similarly. First, they rapidly decline, then their decrease slows down with a downwards trend till epoch 30 and are almost constant till the end.

\begin{remark}\label{rem:dnnMonotonicity}%
	Monotonicity is a well-researched property of DNNs. There are several ways to ensure that an output of a DNN is monotonically increasing. We can add a simple penalty on the gradient of the DNN as a soft-constraint to the loss function as suggested in \cite{Gupta2019} or design the network such that it necessarily outputs monotone increasing numbers as in \cite{Wehenkel2021}. Another easy method would be to use a cumulative sum of positive values of the last layer of the DNN.
	
	However, in our experiments the monotonicity was learned without imposing any hard or soft-constraints by the convolutional network. A pure dense version of this network failed to capture the monotonicity, outputs looked like $x_0=1.0,1.1,\dots,1.9,31,2.0,2.1,\dots$, i.e., sometimes huge values appeared in the otherwise monotonic output.
	Also adding a soft-constraint to the dense architecture could not reach the accuracy of the convolutional network.
\end{remark}

\subsection{CDO calibration}\label{sec:calCDO}
Now, that we have learned how to derive the initial distances to default $x_0$ from the parameters $\rho,\sigma$, such that \eqref{eq:calInner} is satisfied, we can substitute the DNN $f$ into \eqref{eq:calOuter}. This means, in this section we tackle 
\begin{align*}
	\min_{\rho\in [0,1),\sigma\in \Sigma}
		\sum_{j=1}^{J}{
			\left(
				\mathrm{C}_{0,T}^j -
				\mathrm{C}_{0,T}^j(\rho,\sigma,f(\rho,\sigma))
			\right)^2
		}+
		\left(
			\mathrm{I}_{0,T} -
			\mathrm{I}_{0,T}(\rho,\sigma,f(\rho,\sigma))
		\right)^2.
\end{align*}
For this, we need an efficient numerical method to evaluate the tranche $[\aPoint_j,\dPoint_j]$-CDO spreads $\mathrm{C}_{0,T}^j$ and the index $\mathrm{I}_{0,T}$ for which we will use the large basket approximation described in Section \ref{sec:largeBasketApprox}.

\paragraph*{Numerical schemes for the large basket SPDE \eqref{eq:schemeSPDE}.}
Suppose, we are at the $n$-th iteration of \eqref{eq:schemeSPDE}. Let us define the following matrices corresponding to the central finite difference approximation of the space derivatives using the homogeneous grid $\mathbb{X}^d_{a,b}$ in the interval $[a,b]\subseteq \R$ with $d+2$ points.

We will assume zero-boundary condition and define 
\begin{align*}
	\left[\frac{v^{(n)}(t,x_{i+1}) - v^{(n)}(t,x_{i-1})}{2\Delta x}\right]_{i=1,\dots,d}
	&\eqqcolon D^x v_t^{(n)},\\
	\left[\frac{v^{(n)}(t,x_{i+1}) - 2 v^{(n)}(t,x_{i}) + v^{(n)}(t,x_{i-1})}{\left(\Delta x\right)^2}\right]_{i=1,\dots,d}
	&\eqqcolon D^{xx} v_t^{(n)},
\end{align*}
where
\begin{align*}
	D^x \coloneqq \frac{1}{2\Delta x}\mathrm{tridiag}^{d}\left(-1,0,1\right),\
	D^{xx} \coloneqq \frac{1}{\left(\Delta x\right)^2} 
		\mathrm{tridiag}^{d}\left(1,-2,1\right),\
	v_t^{(n)} \coloneqq \left[v^{(n)}(t,x_{i})\right]_{i=1,\dots,d}.
\end{align*}
As described in \cite{KPP2021,KPP2022}, we approximate \eqref{eq:largeBasketSPDE} by
\begin{align*}
	d v_t^{(n)} &= \left(\frac{1}{2}D^{xx}-\beta D^x\right) v_t^{(n)} dt - \sqrt{\rho}\ D^{x} v_t^{(n)} dM_t \eqqcolon \mathbf{B}\ v_t^{(n)}\ dt + \mathbf{A}\ v_t^{(n)}\ dM_t\\
	v_0^{(n)} &= \left[v^{(n-1)}(T_n,x_{i})\1_{x_i> 0}\right]_{i=1,\dots,d}.
\end{align*}
For this, we compute the fundamental solution
\begin{align}
	dX_t = \mathbf{B}\ X_t\ dt + \mathbf{A}\ X_t\ dM_t,
	\quad X_0=I_d \in \R^{d\times d},
	\label{eq:fundamentalSolution}
\end{align} 
and approximate the solution for $v_t^{(n)}$ by
\begin{align}
	v_t^{(n)} \approx X_t\ v_0^{(n)} \approx \exp\left(Y_t^m\right)\ v_0^{(n)}.
	\label{eq:SM}
\end{align}
We recall that \eqref{eq:fundamentalSolution} can be approximated by the $m$-th order Itô-stochastic Magnus expansion 
(cf. \cite{KPP2021,KPP2022}) using the case of constant coefficients.
Let us recall the first two orders of the Magnus expansion formula from \cite{KPP2021,KPP2022} in the case of constant coefficients
\begin{align*}
	Y_t^1 &= \mathbf{B}\ t + \mathbf{A}\ M_t\\
	Y_t^2 &= 
		Y_t^1 
		- \frac{1}{2} \mathbf{A}^2\ t 
		+ \left[\mathbf{B},\mathbf{A}\right] \int_{0}^{t}{M_s ds}
		- \frac{1}{2} \left[\mathbf{B},\mathbf{A}\right]\ t\ M_t,
\end{align*}
where $\exp\left(Y\right)\coloneqq \sum_{k=0}^{\infty}{\frac{1}{k!}Y^k}$ denotes the matrix exponential and $[B,A]\coloneqq BA-AB$ the matrix commutator. We use the order two expansion for our experiments and apply the same performance tips suggested in \cite{KPP2022}, i.e., using sparsity and a special algorithm for the matrix-vector exponentiation on CPU only.

Now, if we also discretize the time-derivative on a homogeneous time grid\newline 
$\mathbb{T}_{n}^L\coloneqq \left\{t_l\coloneqq 0+(l-1) \cdot \Delta t \colon \Delta t =\frac{T_{n+1}}{L-1},\ l=1,\dots,L\right\}$ for the interval $[0,T_{n+1}-T_{n}]$ with $L\in\N$ points, we similarly get the Euler-Maruyama scheme
\begin{align}
	v_{t_{l+1}}^{(n)} =
	v_{t_{l}}^{(n)}+
	\mathbf{B}\ v_{t_{l}}^{(n)}\ \Delta t +
	\mathbf{A}\ v_{t_{l}}^{(n)}\ \Delta M_{t_{l}}, \quad \Delta M_{t_{l}}\coloneqq M_{t_{l+1}}-M_{t_{l}}.
	\label{eq:EM}
\end{align}

\paragraph*{Numerical schemes for the large basket PDE \eqref{eq:schemePDE}.}
We proceed similarly to the previous paragraph for the PDE \eqref{eq:largeBasketPDE} and use the same notation to get  
\begin{align*}
	d u_t^{(n)} = \left(\frac{1-\rho}{2} D^{xx}- \beta D^{x}\right) u_t^{(n)} dt \eqqcolon
	\mathbf{C}\ u_t^{(n)} dt.
\end{align*}
Now, if we want to apply the Magnus expansion to this equation, let us first of all note that the Itô-stochastic Magnus expansion with $\mathbf{A}=0$ coincides with the usual deterministic Magnus expansion.
For a comprehensive overview of the deterministic Magnus expansion, we refer the reader to the excellent work of \cite{Blanes2009}.

We notice that all commutators are zero, therefore the deterministic Magnus expansion is immediately of order infinity and does not introduce another numerical error. We have
the exact solution
\begin{align}
	u_t^{(n)} = \exp\left(\mathbf{C} t\right) u_0^{(n)}.
	\label{eq:DM}
\end{align}
Now, if $T_{n+1}-T_{n}$ is chosen such that they are constant for all $n=1,\dots,N$, then we need to compute the matrix exponential only once, which is very efficient.

Again, if we further discretize the time grid as above, we get the pathwise theta-scheme suggested in \cite{Reisinger2011}:
\begin{align}
	\left(I_d - \theta\ \mathbf{C}\ \Delta t\right) u_{t_{l+1}}^{(n)} = 
	\left(I_d + (1-\theta)\ \mathbf{C}\ \Delta t\right) u_{t_{l}}^{(n)},\quad \theta \in [0,1].
	\label{eq:theta}
\end{align}
Since, $\mathbf{C}$ is constant, this can also be efficiently implemented by first computing the QR- or LR-decomposition in the case $\theta > 0$. In our implementation, we use the Crank-Nicolson scheme ($\theta=\frac{1}{2}$) with Rannacher start-up, i.e., $\theta=1$ for the first four half time-steps.

To evaluate either the deterministic Magnus scheme or the theta-scheme for \eqref{eq:schemePDE} at the shifted points $x_i-\sqrt{\rho}\left(M_{T_{n+1}}-M_{T_n}\right)$, we need to interpolate $u^{(n)}(T_{n+1},x_i)$ in between the grid points $x_i\in \mathbb{X}^d_{a,b}$, since the shifted points usually do not coincide with $\mathbb{X}^d_{a,b}$.
For this, we will use spline interpolation in parallel for all simulations on CPU. 

\paragraph*{Comparison of the schemes.}
Let us now compare the numerical schemes for the large basket approximation and choose based on our findings a suitable candidate for the calibration procedure. 

The maturity $T=5$ and the constant annuity is equal to $\alpha=0.25$ corresponding to quarterly resettlements. The loss given default is fixed to $\LGD=0.6$.

We use $d=201$ points for the space grid $\mathbb{X}_{-10,20}^{201}$ with cut-off $[-10,20]\subset \R$. The stochastic Magnus expansion requires the evaluation of Lebesgue integrals in time for which we set $L^\mathrm{SM}=15$ for the time grid $\mathbb{T}^{L^\mathrm{SM}}_n$. We use the same number of time-steps for the Euler-Maruyama scheme, i.e., $L^\mathrm{EM}=15$. Since the theta-scheme with $\theta=0.5$ is implicit, we only need $L^\mathrm{theta}=5$ points in time for each interval $[T_n,T_{n+1}]$. The deterministic Magnus expansion does not require a time-discretization.

\begin{figure}%
\centering
\includegraphics[width=.5\columnwidth]{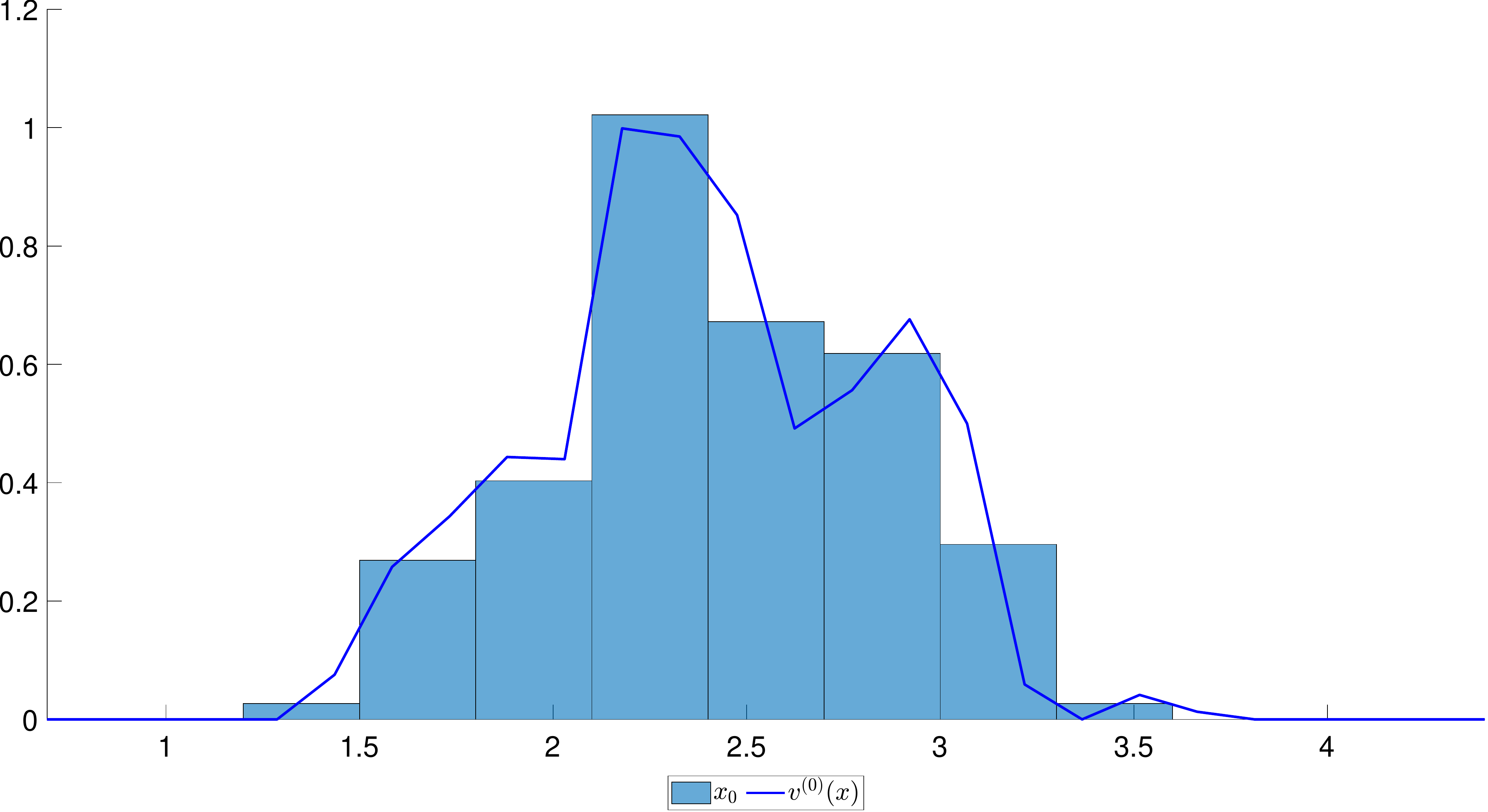}%
\caption{Initial distribution of $x_0$ as a histogram and initial density $v^{(0)}(x)$ used in Table \ref{tab:comparisonSchemes}.}%
\label{fig:x0_26_09_22}%
\end{figure}

In Table \ref{tab:comparisonSchemes}, we show the comparison of the schemes for the same parameters $\sigma=0.0543$, $\rho=0.158$, $r=0.015$ and $x_0$ as in Figure \ref{fig:x0_26_09_22} in terms of the tranche spreads (in bps) with different attachment and detachment points, as well as the corresponding index (in bps) and their computational times.
Figure \ref{fig:x0_26_09_22} shows the histogram of the chosen $x_0\in \R^K$ illustrated by the light blue columns and the corresponding initial density $v^{(0)}(x)$ depicted by a bold dark blue line. The actual values of $x_0$ for this experiment can be found in the \matlab code.

\begin{table}%
\centering
\caption{Comparison of the different numerical schemes for a CDO evaluation with $M=10^5$ simulations of the Brownian motions, $\sigma=0.0543$, $\rho=0.158$, $r=0.015$ and $x_0$ as in Figure \ref{fig:x0_26_09_22}.}
\begin{tabular}{l*{5}{c}}
	Tranche (in bps) & MC 
		& EM \eqref{eq:EM} & Theta \eqref{eq:theta} & SM \eqref{eq:SM} & DM \eqref{eq:DM}\\ 
	\toprule
	$[0,0.03]$ & $ 4447.95 $ & $ 4943.30 $ & $ 4942.32 $ & $ 4929.91 $ & $ 4945.05 $ \\ 
	$[0.03,0.06]$ & $ 1791.54 $ & $ 1890.27 $ & $ 1871.09 $ & $ 1876.80 $ & $ 1877.09 $ \\ 
	$[0.06,0.09]$ & $ 875.99 $ & $ 901.86 $ & $ 887.94 $ & $ 892.95 $ & $ 891.74 $ \\ 
	$[0.09,0.12]$ & $ 439.34 $ & $ 442.01 $ & $ 433.52 $ & $ 436.47 $ & $ 435.66 $ \\ 
	$[0.12,0.22]$ & $ 113.22 $ & $ 109.19 $ & $ 106.92 $ & $ 107.51 $ & $ 107.47 $ \\ 
	$[0.22,1]$ & $ 1.06 $ & $ 0.89 $ & $ 0.87 $ & $ 0.87 $ & $ 0.87 $ \\ 
	\midrule
	Index (in bps) & $ 149.35 $ & $ 153.34 $ & $ 152.24 $ & $ 152.52 $ & $ 152.55 $ \\ 
	\midrule
	c.-time (in s) & $ 2.94 $ & $ 85.34 $ & $ 45.60 $ & $ 78.95 $ & $ 14.07 $ \\ 
\end{tabular}
\label{tab:comparisonSchemes}
\end{table}
The direct Monte-Carlo method has to be understood as a sanity check in this experiment and might not be very accurate, because we use few simulations for this direct approach and also need to simulate $W_t^k$, $k=1,\dots,K$. For the SPDE methods we only need to simulate $M_t$ making them much more stable with respect to the number of simulations. 

We can see that the large basket approximations are very close for all methods and are not too far from the direct Monte-Carlo approach (first column). The Euler-Maruyama scheme was the slowest and the deterministic Magnus expansion the fastest method for the large basket approximation. We also did some experiments on a computer cluster and found that the stochastic Magnus expansion is faster than the theta scheme with 24 CPU cores due to its high parallelizability but it is always slower than its deterministic counterpart with the same number of CPU cores. The usage of a GPU did not improve the computational time for any of the methods.

Since the deterministic Magnus expansion is the fastest for the large basket approximation, we will use it for the calibration in the next section.
\subsection{Calibration Results}\label{sec:calibration}
Now, we are finally able to calibrate \eqref{eq:calOuter} using the pre-trained DNN $f$ and the deterministic Magnus expansion for the large basket approximation. We will fix the number of simulations to $M=10^4$.\footnote{Alternatively, we could use the same technique as in \cite{Reisinger2011}: Starting with few simulations and restarting the calibration with successively increasing simulations, such that the calibration procedure only needs one or two steps for a large amount of simulations. We leave this experiment to the reader, since the code will be publicly available.}

We will use the same parameters and grids for the numerical schemes as in the previous paragraph. In Table \ref{tab:calibrationError}, we present the calibration errors (mean absolute percentage errors) as well as the tranche spreads (in bps) and the index (in bps) at \dateB with fixed short rate $r=0.026$, and in Table \ref{tab:compTimes} the overall computational times (in seconds) of the entire scheme.

\begin{table}[htbp]%
\centering
\caption{Calibration errors using the market data at \dateB with interest rate $r=0.026$.}
\begin{tabular}{lllll}
& $[0,0.03]$ & $[0.03,0.06]$ & $[0.06,0.12]$ & Index \\ 
\toprule 
Market (in bps) & $ 4506.92 $ & $ 1076.74 $ & $ 318.17 $ & $ 100.42 $ \\ 
Calibration (in bps) & $ 4496.10 $ & $ 1268.90 $ & $ 291.52 $ & $ 87.56 $ \\ 
\midrule
Error (in \%) & $ 0.24 $ & $ 15.14 $ & $ 9.14 $ & $ 14.69 $ \\ 
\end{tabular}
\label{tab:calibrationError}
\end{table}

\begin{figure}[htbp]%
\centering
\includegraphics[width=.5\columnwidth]{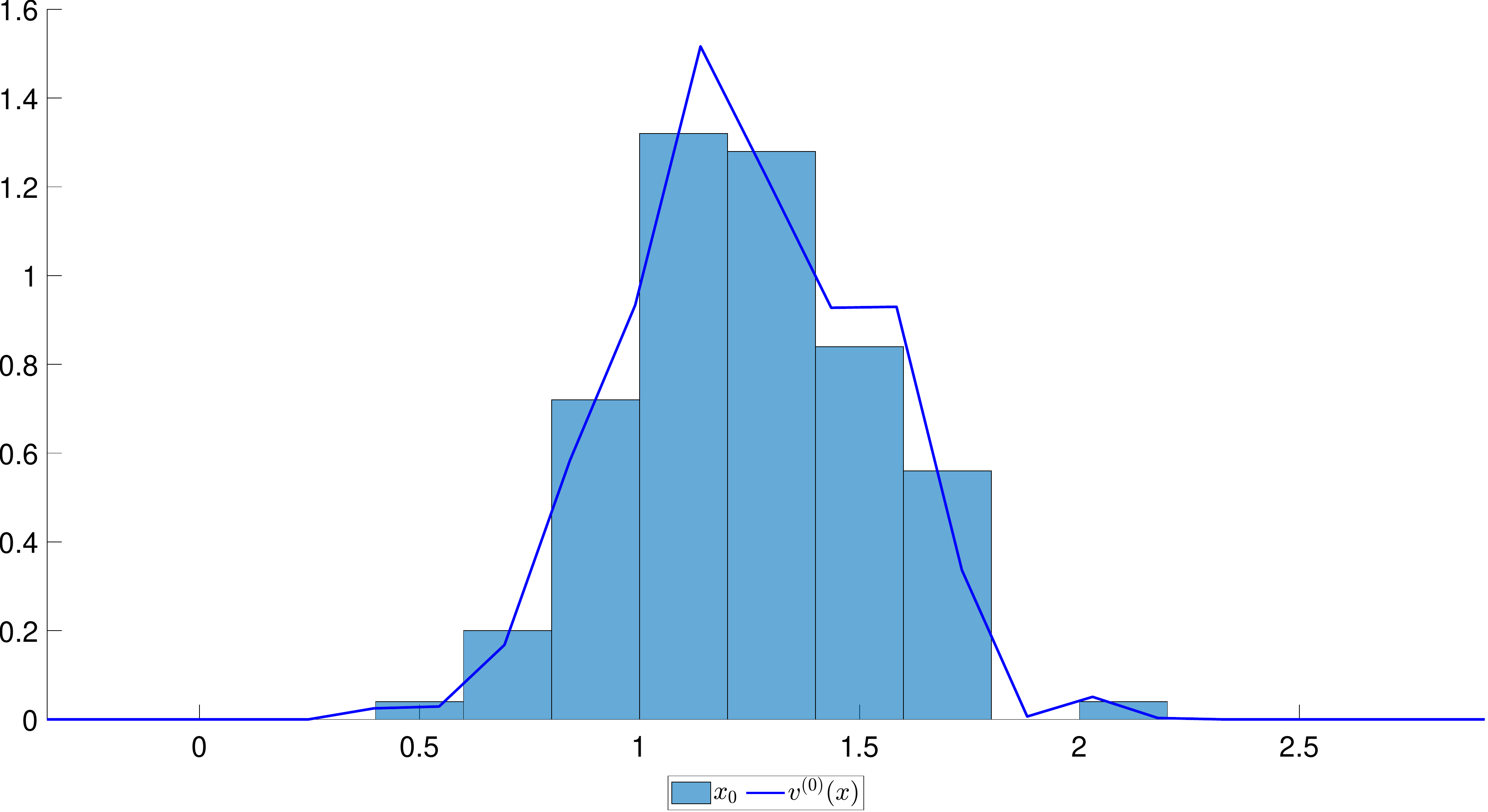}%
\caption{Initial distribution of $x_0$ as a histogram and initial density $v^{(0)}(x)$ used in Table \ref{tab:calibrationError}.}%
\label{fig:x0_05_12_22}%
\end{figure}

We found that small values for the volatility worked best as initial point for the optimization and chose $\sigma_0=0.05$ and $\rho_0=0.5$ as starting points. We also used the same bounds on the parameters as for the neural networks. For the optimization, we use \matlab's \lsqnonlin with Trust-Region-Reflective algorithm. The parameters found by the optimizer were $\sigma=0.0294$ and $\rho=0.2409$ with initial distances of default $x_0$ illustrated in Figure \ref{fig:x0_05_12_22}, analogue to Figure \ref{fig:x0_26_09_22}, and the explicit values can be found in the \matlab code.

\begin{table}[htbp]%
\centering
\caption{Overall computational times using the market data at \dateB.}
\begin{tabular}{lcr}
	Method & Device & Computational times in s\\
	\toprule
	Dataset ($2^{17}$ data points with $M=10^5$) & CPUs & 206.88\\
	Training of $g$ ($30$ epochs each $2^{10} \times 128$) & GPU & 102.77\\
	Training of $f$ ($40$ epochs each $100\times 128$) & GPU & 267.69\\
	Calibration with DM \eqref{eq:DM} & CPUs & 72.21 \\
	\midrule
	Total & & 649.55
\end{tabular}
\label{tab:compTimes}
\end{table}

We can see a good fit to the market data especially for the first tranche and the total computational time is around 11 minutes. This can be further improved by sampling a larger dataset for $g$ and training the DNN for $g$ once.\footnote{Using Julia-Lang with multithreading instead of Python with multiprocessing, Numba and Numpy leads to a further speed-up of roughly a factor 2.5 for generating this dataset. The Julia-code is also publicly available.} We did this experiment with equal performance of DNN $g$ compared to the one presented in this paper. However, as of now the DNN $f$ has to be retrained for different CDS data leading to a computational time of about 6 minutes after the improvement for $g$. Since, the GPU of the machine used in this paper is rather small, we did a test with an
NVIDIA Tesla V100-PCIE-32GB (HBM2), which required only 132.33 seconds for the training of $f$.
\section{Conclusion and Future Research}\label{sec:conclusion}
In this paper, we have shown how to efficiently calibrate a structural CDO model to the available market data by using the large basket approximation. We applied Deep Learning techniques to derive the initial distances to default from CDS quotes of the iTraxx-constituents to remove the pure Monte-Carlo estimation from the calibration Problem \ref{prob:cal1}. Afterwards, we compared four different numerical methods for the large basket SPDE and found that the deterministic Magnus expansion excelled in both accuracy and speed. Then, we calibrated the structural model to the market spreads and index with good accuracy for a model with so few parameters.

It is straightforward to increase the flexibility of this model by adding jumps like \cite{Reisinger2011}, while still using the deterministic Magnus expansion, which should improve the fit to the data further. The stochastic Magnus expansion is unfortunately not available at the moment for jump-diffusions.

The numerical schemes in this paper scale all very well with the number of computing units making this an excellent method for practitioners.

In a future line of research, we would like to study if it is possible to supply the gradients for the optimization algorithm to improve its performance by using a differentiable architecture for the neural networks and the simplicity of the deterministic Magnus expansion.

Moreover, we would like to improve the DNN for $f$ such that it does not have to be re-trained for different CDS quotes. One possibility could involve DeepONets (cf. \cite{Lu2021}).

\appendix

\section*{Declarations}

The authors have no relevant financial or non-financial interests to disclose.

{
\thispagestyle{scrheadings}\ihead{}
\singlespacing
\begin{footnotesize}
\bibliographystyle{acm}
\bibliography{literature.bib}

\begin{thebibliography}{10}

\bibitem{Rutkowsky2004}
{\sc Bielecki Tomasz~R., R.~M.}
\newblock {\em Credit Risk: Modeling, Valuation and Hedging.}
\newblock Springer Finance, 2004.

\bibitem{Blanes2009}
{\sc Blanes, S., Casas, F., Oteo, J.~A., and Ros, J.}
\newblock The {M}agnus expansion and some of its applications.
\newblock {\em Phys. Rep. 470}, 5-6 (2009), 151--238.

\bibitem{Brigo2006}
{\sc Brigo, D., and Mercurio, F.}
\newblock {\em Interest rate models: theory and practice: with smile, inflation
  and credit.}
\newblock Springer, 2006.

\bibitem{Brigo2013}
{\sc Brigo, D., Morini, M., and Pallavicini, A.}
\newblock {\em Counterparty Credit Risk, Collateral and Funding: With Pricing
  Cases For All Asset Classes}.
\newblock The Wiley Finance Series. Wiley, 2013.

\bibitem{Reisinger2011}
{\sc Bujok, K., and Reisinger, C.}
\newblock Numerical valuation of basket credit derivatives in structural
  jump-diffusion models.
\newblock {\em Journal of Computational Finance 15}, 4 (2012), 115--158.

\bibitem{Bush2011Arxiv}
{\sc Bush, N., Hambly, B.~M., Haworth, H., Jin, L., and Reisinger, C.}
\newblock Stochastic evolution equations in portfolio credit modelling with
  applications to exotic credit products, 2011.

\bibitem{CherubiniDellaLunga}
{\sc Cherubini, U., and Lunga, G.}
\newblock {\em Structured Finance: The Object Oriented Approach}.
\newblock The Wiley Finance Series. Wiley, 2007.

\bibitem{Lando2004}
{\sc David, L.}
\newblock {\em Credit Risk Modeling: Theory and Applications.}
\newblock Princeton Series in Finance, 2004.

\bibitem{Duffie2012}
{\sc Duffie~Darrell, S. K.~J.}
\newblock {\em Credit Risk: Pricing, Measurement, and Management.}
\newblock Princeton University Press, 2012.

\bibitem{Reisinger2012}
{\sc Giles, M.~B., and Reisinger, C.}
\newblock Stochastic finite differences and multilevel monte carlo for a class
  of spdes in finance, 2012.

\bibitem{Gupta2019}
{\sc Gupta, A., Shukla, N., Marla, L., Kolbeinsson, A., and Yellepeddi, K.}
\newblock How to incorporate monotonicity in deep networks while preserving
  flexibility?, 2019.

\bibitem{He2022}
{\sc He, W., and Guan, M.}
\newblock Parameter estimation method of option pricing model based on
  convolutional neural network in high frequency financial trading.
\newblock {\em Annals of Operations Research\/} (Mar 2022).

\bibitem{Hu2015}
{\sc Hu, C., Zhang, X., and Gao, Q.}
\newblock Synthetic cdo pricing: the perspective of risk integration.
\newblock {\em Applied Economics 47}, 15 (2015), 1574--1587.

\bibitem{Schonbucher2003}
{\sc J., S.~P.}
\newblock {\em Credit Derivatives Pricing Models: Models, Pricing and
  Implementation.}
\newblock Wiley, 2003.

\bibitem{KM2022b}
{\sc Kamm, K., and Muniz, M.}
\newblock Rating triggers for collateral-inclusive xva via machine learning and
  sdes on lie groups, 2022.

\bibitem{KPP2021}
{\sc Kamm, K., Pagliarani, S., and Pascucci, A.}
\newblock On the stochastic magnus expansion and its application to spdes.
\newblock {\em Journal of Scientific Computing 89}, 3 (Oct 2021), 56.

\bibitem{KPP2022}
{\sc Kamm, K., Pagliarani, S., and Pascucci, A.}
\newblock Numerical solution of kinetic spdes via stochastic magnus expansion,
  2022.

\bibitem{Lu2021}
{\sc Lu, L., Jin, P., Pang, G., Zhang, Z., and Karniadakis, G.~E.}
\newblock Learning nonlinear operators via {DeepONet} based on the universal
  approximation theorem of operators.
\newblock {\em Nature Machine Intelligence 3}, 3 (mar 2021), 218--229.

\bibitem{Wehenkel2021}
{\sc Wehenkel, A., and Louppe, G.}
\newblock Unconstrained monotonic neural networks.

\end{thebibliography}
\end{footnotesize}
}

\end{document}